\documentclass[12pt]{article}

\usepackage{scicite}

\usepackage{times}

\usepackage{graphicx}
\usepackage{dcolumn}
\usepackage{bm}

\usepackage{amsmath}
\usepackage{hyperref}
\usepackage{amsfonts}
\usepackage{subfigure}

\usepackage{mathtools}

\usepackage[dvipsnames]{xcolor}

\newcommand{\half}{\frac{1}{2}}
\newcommand{\la}{\langle}
\newcommand{\ra}{\rangle}

\newcommand{\mudir}{\mu^{(\alpha)}}

\newcommand{\paralpha}{{(\alpha)}}

\newcommand{\invcm}{cm$^{-1}$}

\newenvironment{sciabstract}{%
\begin{quote} \bf}
{\end{quote}}

\title{Near- and long-term quantum algorithmic approaches for vibrational spectroscopy}

\author
{Nicolas P. D. Sawaya,$^{1\ast}$ Francesco Paesani,$^{2,3,4}$ Daniel P. Tabor$^{5\ast}$\\
\\
\normalsize{$^{1}$ Intel Labs, Santa Clara, CA 95054, USA}\\
\normalsize{$^{2}$ Department of Chemistry and Biochemistry, UC San Diego, La Jolla, CA 92093, USA}\\
\normalsize{$^{3}$ Materials Science and Engineering, UC San Diego, La Jolla, CA 92093, USA}\\
\normalsize{$^{4}$ San Diego Supercomputer Center, UC San Diego, La Jolla, CA 92093, USA}\\
\normalsize{$^{5}$ Department of Chemistry,
Texas A\&M University, College Station, TX 77843, USA}\\

\\
\normalsize{$^\ast$Corresponding authors: nicolas.sawaya@intel.com, daniel\_tabor@tamu.edu.}
}

\date{}

\begin{document} 


\baselineskip24pt


\maketitle


\begin{sciabstract}
Determining the vibrational structure of a molecule is central to fundamental applications in several areas, from atmospheric science to catalysis, fuel combustion modeling, biochemical imaging, and astrochemistry.
However, when significant anharmonicity and mode coupling are present, the problem is classically intractable for a molecule of just a few atoms. Here, we outline a set of quantum algorithms for solving the molecular vibrational structure problem for both near- and long-term quantum computers. There are previously unaddressed characteristics of this problem which require approaches distinct from most instances of the commonly studied quantum simulation of electronic structure: many eigenstates are often desired, states of interest are often far from the ground state (requiring methods for ``zooming in'' to some energy window), and transition amplitudes with respect to a non-unitary Hermitian operator must be calculated. We address these hurdles and consider problem instances of four molecular vibrational Hamiltonians. 
Finally and most importantly, we give analytical and numerical results which suggest that, to a given energy precision, a vibrational problem instance will be simulatable on a quantum computer before an electronic structure problem instance. These results imply that more focus in the quantum information community ought to shift toward scientifically and industrially important quantum vibrational problems.
\end{sciabstract}


\section{Introduction}
\label{sec:intro}



To date, the vast majority of chemistry- and materials-related quantum algorithms research has focused on the electronic structure problem \cite{mcardle20_rev,cao19_rev}. 
Given a particular set of nuclear coordinates, the goal is to solve the fermionic (electronic) many-body problem to determine accurate energies.
However, an accurate solution of the electronic structure problem is only one of the current challenges in computational chemistry and materials science. 
There are properties of interest for which the computational bottleneck is not the electronic structure problem, but rather an accurate quantum treatment of the molecular motion \cite{mcardle19_qvibr}.

\begin{figure}
    \centering
    \includegraphics[width=0.6\linewidth]{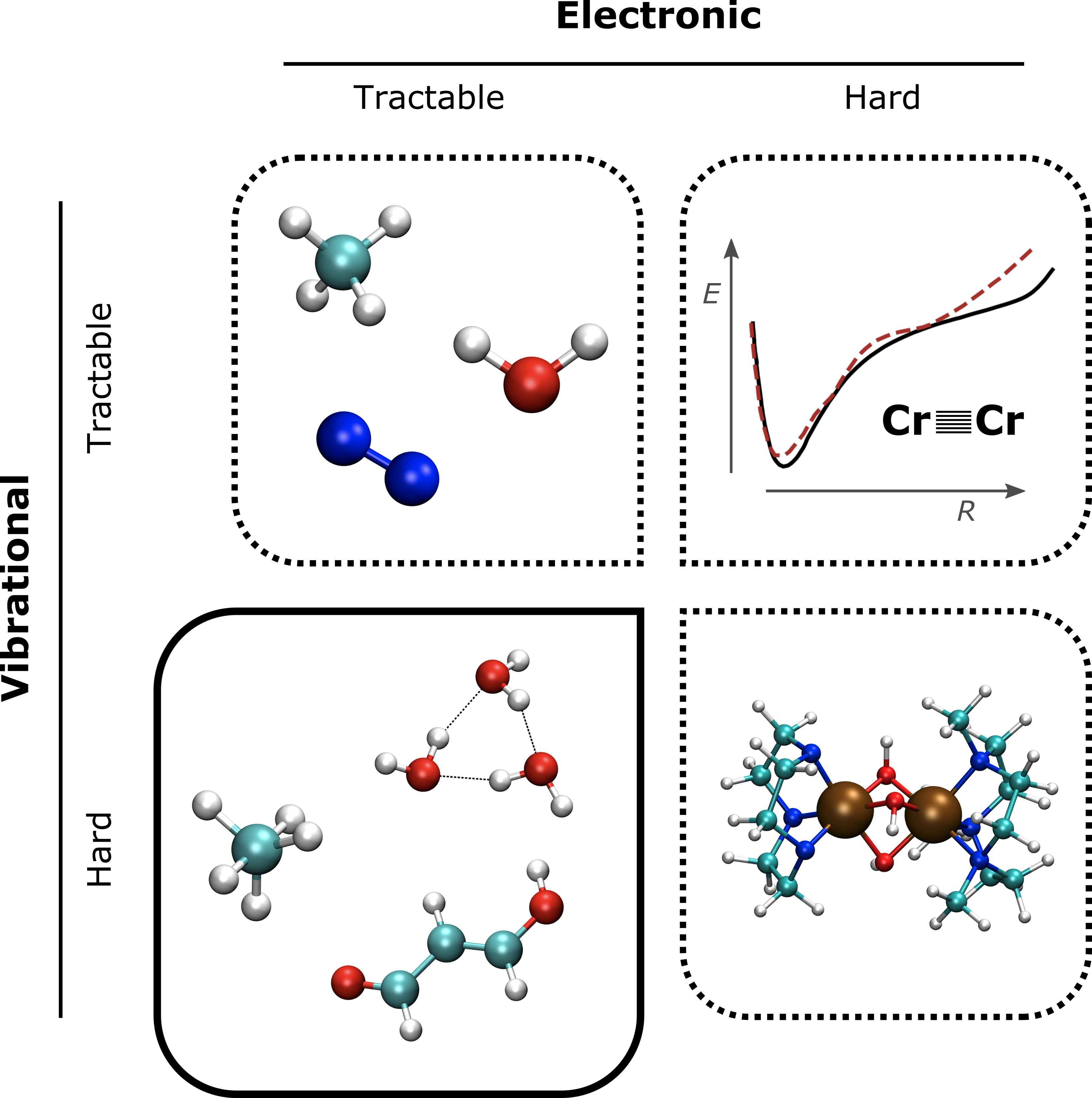}
    \caption{We categorize molecular simulation into four quadrants, depending on whether the electronic structure and vibrational structure of the problem are tractable on a classical computer. This work's focus is the lower-left quadrant---those molecules or complexes for which the potential energy surface can be calculated on a modern classical computer, while the quantum vibrational structure may require a quantum computer. \textit{Upper-left}: small rigid molecules methane, water, and nitrogen. \textit{Upper-right}: the chromium dimer, for which state-of-the-art electronic structure methods (dashed) cannot match experiment (solid) \cite{sheng_cr2}. \textit{Lower-right}: A chromium-containing large molecule with non-rigid ligands. \textit{Lower-left}: the prototypical ``fluxional'' molecule CH$_5^+$, the water trimer which contains non-bonded interactions, and malondialdehyde which contains migratory hydrogen atoms.}
    \label{fig:quadrants}
\end{figure}

One such area is the calculation of vibrational spectra, as there is a large subset of molecules for which the electronic structure problem \textit{is} classically tractable to subchemical accuracy
while the quantum vibrational problem is not (see Figure \ref{fig:quadrants}). 
This is true for small molecules and clusters in several areas of spectroscopy: infrared spectra, Raman spectra, vibronic spectra, and ultrafast vibrational spectra, to name just a few \cite{bernath05,mukamel99,wilson80}. 
Roughly speaking, the electronic structure problem is harder the larger the molecule is and the more electron correlation that is present (due to \textit{e.g.} transition metal elements). 
The vibrational structure problem, on the other hand, is hard for non-rigid or ``fluxional'' molecules as well as non-covalent complexes \cite{jankowski12_vdw} such as aqueous clusters \cite{brown17_watermelt,Yang:2019waterscience,riera18_metalwater,bajaj19_h2o_i}. 


Even qualitatively correct vibrational spectra often require a rigorous quantum treatment because Fermi resonances, association bands, and other resonance effects can result from small coupling terms \cite{bernath05,vazquez07_fermi,sibert14_fermires,mccoy12_associationband}. 
The ability to calculate vibrational structure has many applications including fuel combustion \cite{harding17_combust}, atmospheric science \cite{hou13_atmosci}, astrochemistry \cite{barone15_astrochem,mcguire18_astrochem},
and fundamental experiments in chemical physics \cite{ospelkaus10_polar_rovibr,Ribeiro:2018polaritons}.
Other than vibrational spectra, problems that lie in the lower-left quadrant of Figure \ref{fig:quadrants} include low-temperature thermodynamic calculations of some bulk solids \cite{kapil19_solids} and quantum liquids \cite{qliquid_book}.

Previous quantum computational studies in this area include analog quantum algorithms for quantum vibrations \cite{huh15,sparrow18_vibrdyn,mahesh14_fcnmr,teplukhin19_vibr_anneal,macdonell_12},
digital quantum algorithms for finding vibrational states and/or overlaps \cite{mcardle19_qvibr,sawaya19_vibronic,ollitrault20_reiher_qvibr},
and approaches for which vibrational degrees of freedom are coupled to other systems \cite{veis2016quantum,macridin18a,magann20_qcontrol,ollitrault20_nonadiab}.  
The present work is distinct in several ways as outlined below.

In this work we present algorithms for calculating vibrational spectra on both near- and long-term hardware, focusing on vibrational infrared spectra. One of our contributions is identifying certain essential components that are required for this problem class, drawing attention to the algorithmic objectives that would not appear in most problems that involve Hamiltonian simulation.
Some aims of this class of problems are significantly different from the electronic structure problem, which we introduce to the reader and demonstrate how to overcome. 
Finally and most importantly, we compare the problem's complexity characteristics to the electronic structure problem. The results suggest that, for a given precision, a vibrational structure problem instance will probably achieve quantum advantage before an electronic one.

From a Hamiltonian simulation perspective, we note three conceptual differences between the nature of the typical electronic structure problem instance and that of vibrational spectroscopy, the first two of which to our knowledge have not been identified previously in this context.
First, when it is said that the ``spectrum'' is being calculated in molecular electronic structure, this normally refers to at most a handful of the lowest-lying electronic states. 
In contrast, in a vibrational problem one is almost always interested in many states, which may indeed be far from the ground state (excited states greater than 100 are often of interest).

The second conceptual difference is that one is often interested in calculating both vibrational energies and transition intensities, which necessitates calculating the transition amplitudes with respect to a non-unitary coordinate-dependent operator. 
Though such transition amplitudes are applicable to electronic structure in some important areas \cite{kosugi20_linresp_chargespin,cai20_qmolecresp}, 
their inclusion has not been the norm in the context of quantum algorithms. 
Third, the problem requires that bosons (vibrations) instead of fermions be encoded into the quantum device, a topic that has been previously explored \cite{somma03_inclboson,mcardle19_qvibr,sawaya20_dlev,sawaya20_connect}.

\section{Theory}
\label{sec:theory}


%
In arbitrary internal coordinates $\vec s$ (with corresponding momenta $\vec m$), the Hamiltonian for $M$ vibrations in general form is written (with $\hbar=1$)
\begin{equation}\label{eq:ham_general}
H = \half \sum g_{ij} m_i m_j + V(\vec s).
\end{equation}
where $g_{ij}$ is the coupling between momenta (vanishing for $i \neq j$ under normal or Cartesian coordinates) and $V(\vec s)$ is the potential energy term. 
In the harmonic approximation, one may diagonalize the Hessian matrix at the equilibrium position, leading to a simplified approximate expression with uncoupled coordinates,

\begin{equation}\label{eq:ham_harm}
H_{\textrm{harm}} = \half \sum_i^M \omega_{i}(q_i^2 + p_i^2),
\end{equation}
\noindent where $i$ denotes the vibrational mode, $M$ is the total number of modes, 
$q$ and $p$ are respectively the bosonic position and momentum operators, and $\omega_i$ is the energy of mode $i$. 
It is trivial to find eigenvalue-eigenfunction pairs for equation \eqref{eq:ham_harm} on a classical computer, since excited states in the Harmonic approximation are product states of separate modes.

Expression \eqref{eq:ham_harm} can be systematically improved by including higher order anharmonic terms,
\begin{equation}\label{eq:ham_aharm}
H_{\textrm{anharm}} = \half \sum_i^M \omega_{i}(q_i^2 + p_i^2) + \sum_{\{ijk\}} h_{ijk} q_iq_jq_k   + \sum_{\{ijkl\}} h_{ijkl} q_iq_jq_kq_l + \cdots,
\end{equation}
where the index ordering is irrelevant and $h_{ijk\cdots}=0$ if all indices are distinct. 
Computational difficulties arise when these higher-order terms are included, due to both the deviation from harmonicity and the coupling between modes. 

Even for a molecule of 5 to 8 atoms, the complete inclusion of anharmonic effects can be computationally prohibitive. 
Though various forms of perturbation theory and dimensionality reduction sometimes yield good results, one must often resort to exact diagonalization of the whole Hilbert space or similarly expensive methods \cite{Bowman:2008MolPhysReview,Light:1987:JCPDVR,Sibert:1988VV,Wang:2008CH5,vanderAvoird:2011:JCPH2OH2,Lin:2015}. 
We note that we are not constrained to use equation \eqref{eq:ham_aharm} but may choose any convenient coordinate system---it will often be the case that choosing a specialized coordinate system allows one to use a lower-order series expansion \cite{Simons:1973_SPF_coords,Zuniga:2005_optcoord,Bulik:2017}. 







In order to make our discussion concrete, we consider infrared spectroscopy, though similar mathematical methods would be used for other experiments such as Raman, microwave, or ultrafast multidimensional vibrational spectroscopy \cite{bernath05,mukamel99}.

The dipole moment operator is necessary for simulation of light-matter interaction, \textit{e.g.} for calculating transition intensities. 
It is denoted $\mudir$ where $\alpha \in \{x,y,z\}$ is a Cartesian direction. 
As $\mudir$ is coordinate-dependent, it is associated with a dipole moment surface (DMS), which may be expanded in a power series,

\begin{equation}\label{eq:mu_expand}
\begin{split}
\mudir &= \mudir_0 
+ \sum_i^M \left. \frac{\partial \mudir}{q_i}\right|_{q_i=0} q_i 
+ \frac{1}{2} \sum_{ij}^M \left.\frac{\partial^2 \mudir}{q_iq_j}\right|_{q_i,q_j=0} q_iq_j + \cdots \\
&= \mudir_0 + \sum_i^M m_{i} q_i + \sum_{ij}^M m_{ij} q_iq_j + \cdots
\end{split}
\end{equation}
The objective is to calculate
\begin{equation}\label{eq:spec_zerotemp}
f(\omega) = \sum_{\alpha} \sum_j |\la 0 | \mudir | j \ra|^2 \mathcal L(\omega_j-\omega)
\end{equation}
%
%
%
%
where $|0\ra$ is the initial eigenstate (ground state when beginning from zero temperature), $\mu^{(\alpha)}$ is the dipole operator for Cartesian direction $\alpha$, and $\mathcal L(\omega)$ is a line shape function, approximated as a delta function when one does not consider broadening effects. 
Though in this work we consider transitions from the ground state, the initial state of interest is often a Gibbs state (\textit{i.e.} thermal state). 
Existing quantum algorithms for thermal state preparation \cite{poulin09_thermal,riera12_thermal,motta19_qite,chowdhury20_thermal,wang20_thermal} 
may be used in conjunction with the approaches summarized here. 
Note that expression \eqref{eq:spec_zerotemp} is mathematically similar to what is used to calculate Franck-Condon factors \cite{huh15,sawaya19_vibronic}, 
where $\mu^{(\alpha)}=I$ and a different Hamiltonian is used.

Both the Hamiltonian and the dipole operator may be mapped to a qubit-based Hamiltonian using the bosonic commutation relations, where a practical choice is to use the Pauli operator basis:

\begin{equation}\label{eq:ham_to_pauli}
H_\textrm{anharm} \mapsto \sum_k^{N_P} a_k P_k = \sum_k a_k \bigotimes_g^{N_q} \sigma_{gk},
\end{equation}

\begin{equation}\label{eq:mu_to_pauli}
\mudir \mapsto \sum_k^{N_R^\paralpha} b_{k\alpha} R_{k\alpha} = \sum_k b_{k\alpha} \bigotimes_g^{N_q} \sigma_{gk\alpha},
\end{equation}
where $g$ labels the qubit, $\sigma_g \in  \{I,X,Y,Z\}$ is the identity or a Pauli operator, $N_P$ ($N_R^\paralpha$) is the number of Pauli strings in the encoded operator, and $N_q$ is the number of qubits. 
Several encodings for performing this mapping have been discussed previously \cite{somma03_inclboson,veis2016quantum,mcardle19_qvibr,sawaya20_dlev}, with evidence that the Gray code offers reasonable resource trade-offs \cite{sawaya20_dlev}.


A near-term algorithm for the vibrational spectroscopy problem requires several elements: (a) mapping of bosons to qubits, (b) finding unitaries $U_i$ to produce eigenstates, (c) determining state overlaps $|\la \psi_i|\psi_j \ra|^2$, (d) calculating transition amplitudes with respect to a non-unitary Hermitian operator, and (e) efficiently finding eigenstates far above the ground state.
We first present the noisy intermediate-scale quantum (NISQ) approach for these problem requirements, before briefly summarizing a long-term approach that addresses all algorithmic requirements.






\subsection{Near-term algorithms.}

Using near-term quantum hardware, ground and excited states may be found using previously published variational methods \cite{higgott19_vqd,jones19_discovspectra,motta19_qite,mcclean16_njp,wang94,ollitrault19_eom}.  
For a given vibrational eigenstate $|\psi_j\ra$, a variational method used with a classical optimizer will lead to a circuit unitary $U_j$ yielding $U_j|0\ra = |\psi_j\ra$. 
Additionally, expression \eqref{eq:spec_zerotemp} requires a method for calculating state overlaps $|\la \psi_i|\psi_j \ra|^2$, for which quantum subroutines are summarized in the Supplementary Information (SI).





In order to calculate arbitrary transition amplitudes $|\la\psi_i| A|\psi_j \ra|^2$ on near-term hardware, a naive approach would require an efficient method for considering cross-terms such as $\la\psi_i|0\ra$ or $\la\psi_i|(\Pi_k \sigma_k)|0\ra$ (where $(\Pi_k \sigma_k)$ is a Pauli string), in addition to their absolute values squared. 
This is a nontrivial task, since quantum computers naturally output overlaps squared. 
Though inner products may be calculated with so-called Hadamard tests that require a substantially increased circuit depth if only one- and two-qubit gates are allowed \cite{mitarai19_newhadamtest}, 
a much shorter-depth method was recently found \cite{ibe20_transition} for calculating transition amplitudes of arbitrary operators. 

Tailoring the latter work to vibrational spectroscopy, the procedure is to use an additional set of unitaries,
\begin{equation}\label{eq:V_kl}
V_{kl,\pm}^\paralpha = \half (I \pm R_{k\alpha}) (I \pm R_{l\alpha}) = e^{\pm R_{k\alpha} \pi/4} e^{\pm R_{l\alpha} \pi/4}
\end{equation}
for all $l,k<N_R^{(\alpha)}$. 
One then proceeds to reproduce $|\la i | \mudir | j \ra|^2$ from many measurements on the circuit set $U_i^\dag V_{kl,\pm} U_j|0\ra$ (see SI). 
Thus one increases the depth of two state preparation circuits by a small constant factor 
and collects measurement statistics from a quadratic (in $N_p$) number of circuits. 
This procedure is performed for every eigenenergy for which one wishes to calculate the transition amplitude. The algorithm requires a larger total number of measurements, but the fact that it increases circuit depth by only a small factor makes it ideal for near-term hardware.







%
\subsection{Spectral window focusing.}
%
%
Not only are we often interested in many vibrational eigenstates---it is also often the case that one is concerned \textit{only} with high-lying excited states (for instance the 100th excited state and above).
This may be the case when: part of a spectrum is blocked by background noise; an astronomical telescope is able to read only part of the infrared spectrum; or only a specific band is technologically relevant.
%
%
Additionally, different regions of the spectrum contain different information. 
For example, in water clusters relevant to atmospheric chemistry, the OH stretches (3800 to 3000 \invcm) report on local environments while the low-frequency region reports on collective motion \cite{perakis16_waterrev}. 
%
Therefore it may be a waste of computational effort to find eigenstates outside the energy window of interest.

We point this out because it means the \textit{goal} of the vibrational spectrum problem often differs from other Hamiltonian simulation problems, leading to important considerations in algorithm design that have not been widely investigated.


The notable consequence is that most hereto proposed near-term algorithms for excited states are not always viable.
%
This is because in their canonical forms, most existing near-term approaches \cite{higgott19_vqd,mclean17_qse,santagati18_waves,jones19_discovspectra,motta19_qite,ollitrault19_eom}
require one to find the low-energy eigenvector subspace as a way to build up to the desired excited state. In other words, if one is not interested in the first $W$ excited states, then the larger $W$ is the more one would like to largely avoid the costly determination of the subspace spanned by $\{\psi_1,\psi_2,\cdots,\psi_W\}$, where $\psi_m$ is the m\textit{th} excited state. 


We highlight one possible (previously proposed) near-term algorithmic solution for determining high-lying excited states, that may be used in conjunction with the previously mentioned methods. 
This is to use the folded spectrum method \cite{mcclean16_njp,wang94}, 
which easily allows one to select an energy neighborhood. 
The folded Hamiltonian is defined as
\begin{equation}
H_{fold} = (H - \zeta I)^2
\end{equation}
where $\zeta$ is an arbitrary constant. 
The lowest eigenstates of $H_{fold}$ are those eigenstates of $H$ which are closest in energy to $\zeta$. 
This approach quadratically increases the number of Pauli terms in the effective Hamiltonian, allowing one to ``zoom in'' on an arbitrary portion of the spectrum. We do not rule out more efficient methods for high-lying excited states.



\subsection{Utility of incomplete spectra.}
When using variational algorithms and NISQ hardware to determine portions of the spectra, one may be able to calculate only incomplete spectra. 
This is due to the nature of many hybrid quantum-classical algorithms; it is usually not possible to guarantee that all eigenstates in a given energy region have been found. 
Hence it is important to note that even a spectrum with missing peaks is often useful. 
First, one may be interested in only a few specific spectral features in the region, in which case one may focus efforts converging to those specific transitions.
Second, and perhaps more importantly, the goal is often to determine whether a candidate molecule matches an experimental result. 
If some spectral features in the computed spectrum of the candidate molecule are not present in the experimental spectrum, then the candidate molecule may be removed from consideration. 



\subsection{Long-term hardware.}
%
%
A long-term fully error-corrected solution to these problems is to use quantum phase estimation (QPE) \cite{abrams99_qpe} to produce a probabilistic set of measurements. 
This approach has been discussed previously \cite{wecker15_correlec,sawaya19_vibronic,Roggero19_linresp}, first by Wecker and co-workers \cite{wecker15_correlec}. 
In the more typical use of QPE, one first attempts to prepare a state with as much overlap as possible with a particular eigenstate, \textit{e.g.} the ground state. In the context of the this work, QPE is instead used in a way that allows one to calculate a full response spectrum \cite{wecker15_correlec,sawaya19_vibronic,Roggero19_linresp}, \textit{i.e.} determining the non-negligible values $| \la \eta_0 | \hat A | \psi_i \ra |^2$, where $\hat A |\eta_0\ra$ is not necessarily an eigenstate of the Hamiltonian but the $\{|\psi_i\ra\}$ are eigenstates. 
First consider the case of $\hat A = \hat I$. One runs the same QPE algorithm, but sets the initial state to $|\eta_0\ra$, which is in general not an eigenstate, such that $|\eta_0\ra = \sum_i c_i |\psi_i\ra$. After running QPE, one is left with a superposition of eigenstate-eigenphase pairs, as shown in the expression

\begin{equation}\label{eq:apx-qpe-approach}
\sum c_i |\psi_i\ra|0\ra \xrightarrow{\textrm{QPE}} \sum c_i |\psi_i\ra|\tilde\phi_i\ra.
\end{equation}

In contrast to the standard use of QPE, in this case we are interested in more than just one eigenstate. 
The algorithm proceeds as follows. One performs many repetitions of the circuit, measuring register $E$ after each run, yielding a phase $\tilde \phi_i$. From many measurements one then composes a histogram where each bin is an $N_E$-bit value $\tilde \phi_i$. This histogram is the desired response spectrum with resolution determined by $N_E$, and the process terminates once the histogram has converged.

One advantage of this method is that it effectively combines many eigenstates into a single measurement. This is especially useful for vibrational spectra, where one is often interested in more than just a few eigenstates of the spectrum. 
For a particular $N_E$, there is a subset of eigenstates $\mathcal D_j$ $= \{ \psi_{j1}, \psi_{j2} \cdots \}$ all of which yield $\tilde \phi_j$. Hence if the measurement yields $\tilde \phi_j$, this means register $S$ has collapsed to the superposition $\mathcal N \sum_{k \in \mathcal D_j} c_{k} |\psi_k\ra$, where $\mathcal N$ is a normalization constant. The beneficial result is that the probabilities of many nearby eigenenergies are combined, and the number of required measurements is dependent on $N_E$ but \textit{independent} of the size of the problem Hamiltonian.


As discussed, vibrational (\textit{e.g.} infrared) spectroscopy requires calculating the action of an arbitrary non-unitary operator $\mudir$ on a prepared state. Roggero \textit{et al.} \cite{Roggero19_linresp} solved the problem of linear response with respect to a non-unitary operator. 
After adding one ancilla qubit, one may apply the operator

\begin{equation}
U_{\mu,\alpha,\gamma} = 
\begin{pmatrix}
\cos \gamma \mudir & -\sin \gamma \mudir \\
\sin \gamma \mudir & \cos \gamma \mudir \\
\end{pmatrix}
\end{equation}
to an arbitrary state $|\psi\ra$, which will yield $\mathcal N \mudir|\psi\ra$ with probability $P_{success} = \la\psi|\sin(\gamma\mudir)|\psi\ra$, where $\mathcal N=||\mudir|\psi\ra||^{-1}$ is a normalization constant.
This unitary probabilistically produces the desired state $|\Phi_0^{(\alpha)}\ra \equiv $ $\hat \mu^{(\alpha)} |\eta_0\ra/ \lVert \hat \mu^{(\alpha)} |\eta_0\ra \rVert$. If the ancilla is measured to be $|0\ra$ ($|1\ra$) then the state preparation has succeeded (failed). The remainder of the algorithm then proceeds as in the $\hat A = \hat I$ case, with $|\eta_0\ra$ replaced by $|\Phi_0^{(\alpha)}\ra$.

Finally, we posit that there are promising strategies for ``spectral window focusing'' in long-term hardware as well. 
For the QPE-based method, the goal would be to make the histogram measurements fall primarily within a particular energy window, as measurements outside the window are not of interest.
In principle one may use amplitude amplification methods \cite{grover96,grover98_ampl} to boost the probability of the desired eigenenergy window. 
The result is that fewer measurements would be required to produce the histogram in the energy window of interest, at the cost of an increase in circuit depth. 
We leave a full description to future work.

\section{Comparison to electronic structure}
\label{sec:vibr_vs_elec}

The first physics simulation to achieve quantum advantage is likely to be a nearest-neighbor toy model such as an Ising model \cite{childs18_pnas}, because only $\mathcal O(N)$ two-body interactions are present. 
But it is important to consider what will be the first real-world non-toy simulation to show quantum advantage. 
Here we argue that, to a given energy precision, the first such simulation of a molecule is more likely to be a vibrational problem instance than an electronic one.
%
Note that the electronic structure Hamiltonian may be written as
\begin{equation}\label{eq:elecham}
H_{\textrm{ES}} = \sum h_{ij} a_i^\dag a_j + \sum h_{ijkl} a_i^\dag a_j^\dag a_k a_l,
\end{equation}
where $a_i^\dag$ and $a_i$ are fermionic creation/annihilation operators for the $i$th orbital and coefficients $h_{ij}$ and $h_{ijkl}$ are determined by calculating overlap integrals. 
2- and 4-body terms are present, and in real molecules one effectively sees nearly all-to-all connectivity between the electronic orbitals. 
For simplicity we write equation \eqref{eq:elecham} without spin constraints (see SI), though our results do account for spin degrees of freedom.

The quantum resources required for the vibrational problem depend on the order of the expansion needed for sufficient precision in equation \eqref{eq:ham_aharm}. 
We observe that early molecular targets for quantum computing ought to be those for which (a) classical computational approaches (\textit{e.g.} perturbation theory) fail and (b) the highest relevant order is 4 or less.
Both requirements are likely to hold for a substantial set of molecules \cite{fortenberry13_qff,Sibert:2019_Review}.

A third-order Hamiltonian has four types of terms: $p_i^2$, $q_i^2$, $q_i^3$,  $q_i^2q_j$.
A fourth-order Hamiltonian has 8 types with the inclusion of $q_i^4$, $q_i^3q_j$, $q_i^2q_j^2$, and $q_i^2q_jq_k$. 
In the SI we give Pauli operator counts for each of these 8 term types, for $d=4$ (2 qubits) and $d=8$ (3 qubits).

A key insight is that fourth-order vibrational Hamiltonians scale at most as $\mathcal O(M^3)$ in the number of modes $M$. However, depending on the choice of coordinate system \cite{Simons:1973_SPF_coords,Zuniga:2005_optcoord,Bulik:2017,Sibert:2019_Review} 
it is often possible to exclude three-body interactions, leading to a scaling of $\mathcal O(M^2)$ terms. This scaling is more favorable than molecular electronic structure, for which near-term implementations would require approximately $\mathcal O(N^4)$ Hamiltonian terms in the number of orbitals $N$. Notably, it is more common to see sparser interactions in vibrational problems than in electronic structure problems, because it is often the case that some vibrational modes show negligible coupling to the other modes---this implies $\mathcal O(M^2)$ may often be an overestimate.

Our case hinges on the notion that a Hamiltonian with more terms and with higher Pauli lengths is likely to require more resources to simulate, regardless of whether one is using near- or long-term hardware.  
The asymptotic scaling on its own does not prove the argument though. One must still investigate (a) whether the pre-factor to the vibrational Hamiltonians' term count is sufficiently small for lower qubit counts, (b) the distributions of lengths of the Pauli strings, and (c) Hamiltonian magnitudes, all of which we study below.

Notably, there has been extensive recent progress in reducing the asymptotic scaling of quantum algorithms for electronic structure \cite{babbush18_lowdepthmats,babbush18_linearT,berry19,vonburg20_catalysis}. 
However, due to these newer algorithms' need for a larger basis set and/or an increase in the number of required qubits, these methods are not amenable to accurate molecular simulation for qubit counts below 100 \cite{vonburg20_catalysis}. 
Even if one assumes that error-corrected hardware is required for solving any instance of both the vibrational and electronic problems, it remains likely that the subset of vibrational problems with $\mathcal O(M^2)$ or $\mathcal O(M^3)$ Pauli terms will be solvable before any electronic structure instance, based on our analysis. Caveats and important further justification of our methods of comparison can be found in the SI.

%



\begin{figure}
    \centering
    \includegraphics[width=1.0\linewidth]{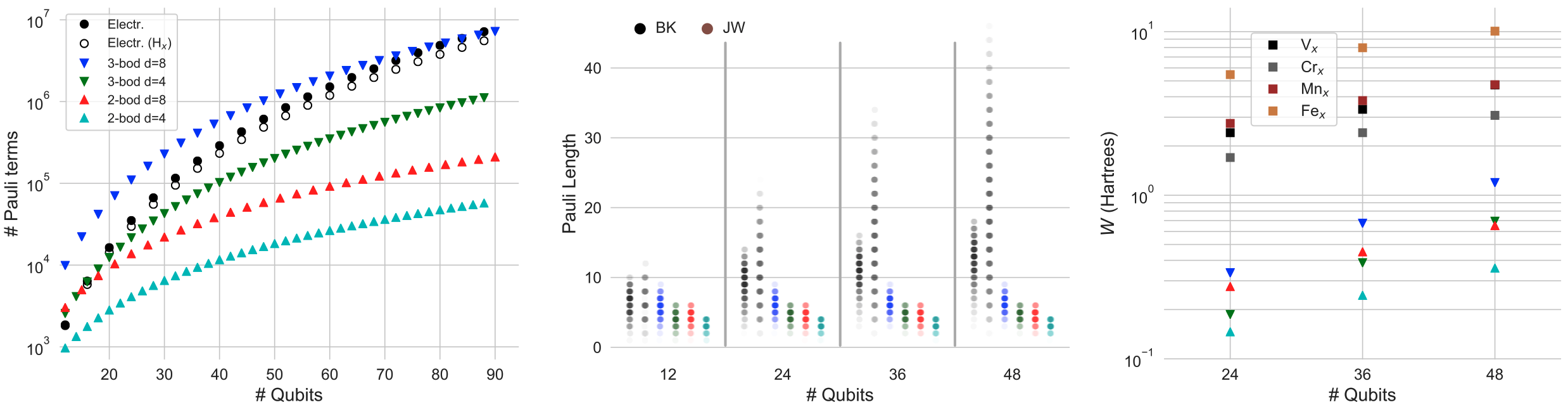}
    \caption{\textit{Left:} Number of Pauli strings in the qubit Hamiltonian versus the number of qubits, for the standard electronic structure problem and some classes of vibrational structure. Filled circles represent analytical results and unfilled circles represent numerical results from a collection of hydrogen atoms. Triangles represent analytical results for fourth-order vibrational Hamiltonians, where \textit{e.g.} \textit{3-bod $d$=8} signifies a truncation of 8 vibrational levels where at most three-body terms are included. \textit{Center:} Probability distributions of Pauli lengths in six Hamiltonian classes. BK and JW respectively denote the Bravyi-Kitaev and Jordan-Wigner mappings for the electronic structure problem. \textit{Right:} The magnitude $W$, which is related to the Frobenius matrix norm, for various Hamiltonians. $W$ should be compared only between Hamiltonians of the same size. The squares signify model electronic structure Hamiltonians of V$_x$, Cr$_x$, Mn$_x$, and Fe$_x$, for $x \in \{2,3,4\}$. Pessimistically large couplings are used for the vibrational Hamiltonians (see SI). All vibrational results use fourth-order Hamiltonians.}
    \label{fig:pauli_elec_vibr}
\end{figure}



\section{Results}
\label{sec:res}


\textbf{Proxies for quantum resource comparisons}. All our vibrational data is for fourth-order Hamiltonians, where we use bosonic truncations $d=$ 4 or 8 and have considered both the exclusion and inclusion of 3-body terms $q_i^2q_jq_k$. 
Though the electronic structure problem scales as $\mathcal O(N^4)$ against the $\mathcal O(M^3)$ or $\mathcal O(M^2)$ vibrational scaling of \textit{some} molecules, the latter always have a larger constant factor (left, Figure \ref{fig:pauli_elec_vibr}). 
However, for the majority of cases considered here, these results show that the electronic structure Hamiltonians contain more terms for qubit counts great than $\sim$20. 

For simplicity, we consider only cases in which all modes have equal $d$. 
In reality, each mode would require a different truncation, meaning that the number of Pauli strings would lie in between the plotted trends. 
For the electronic problem instances, the analytical results (filled circles) are comparable to the numerical results (open circles) obtained from 3D arrays of hydrogen atoms (see SI). 
Note that the number of Pauli terms is equal for the Jordan-Wigner (JW) \cite{jw28} 
and Bravyi-Kitaev (BK) \cite{bravyi02}
encodings, though their length distributions are unequal.


The center panel of Figure \ref{fig:pauli_elec_vibr} shows the distribution of Pauli lengths, another important indicator of a problem Hamiltonian's complexity. 
For the subset of vibrational problem instances considered (fourth-order Hamiltonians with truncations of $d\leq8$), vibrational problems are more local than electronic problems, even for low qubit counts and when compared against the logarithmically scaling BK mapping.

Another factor determining simulation complexity is the magnitude of the Hamiltonian. There are different matrix norms used in quantum algorithm analysis, and resource bounds are usually derived in terms of both a norm and the desired precision, among other considerations \cite{berry15_hamsimoptimal,berry15_hamsimtaylor,childs18_pnas,low17_qsp,childs19_theorytrott}. In addition, the number of measurements for VQE depends on the magnitude of the terms \cite{mcclean16_njp,gonthier20_meas}, which is closely related to our expression below. For a simple and easily computed comparison of Hamiltonian magnitudes, we use the quantity
\begin{equation}\label{eq:Wnorm}
W = \sqrt{ \sum_{k \neq I} a_k^2 },
\end{equation}
where $k \neq I$ signifies that the coefficient preceding the identity operator is excluded. Notably, $W 2^{N_q/2}$ is an upper bound to the Frobenius norm. $W$ should be used for comparisons only between Hamiltonians on the same number of qubits.

We constructed minimal-basis electronic structure model Hamiltonians for transition metals V$_x$, Cr$_x$, Mn$_x$, and Fe$_x$, where $x =$ 2, 3, and 4 correspond to 24, 36, and 48 qubits, respectively (see SI). These were meant to provide typical order-of-magnitude matrices for transition metal elements. We constructed vibrational Hamiltonians with deliberately pessimistic couplings. Harmonic values were evenly spaced between 1000 and 4000 \invcm, every third-order vibrational term was set to 400 \invcm, and every fourth-order term was set to 40 \invcm.

Despite the pessimistically complex vibrational Hamiltonians, $W$ values for the two-body Hamiltonian are close to an order of magnitude smaller than the four types of electronic model Hamiltonians, while $W$ for the three-body vibrational Hamiltonians remain several times smaller (right panel of Figure \ref{fig:pauli_elec_vibr}). 

Additionally, the lower locality of the vibrational Hamiltonians imply that $W$, if used as an approximate proxy for simulation complexity, further underestimates the difference between vibrational and electronic problems. This is because some long-term Hamiltonian algorithms \cite{childs19_theorytrott} scale with locality (center panel of Figure \ref{fig:pauli_elec_vibr}), and near-term VQE requires fewer measurements with a more local Hamiltonian \cite{mcclean16_njp,gonthier20_meas}.

We note again that the required quantum resources are precision dependent. For both vibrational and electronic problems, different applications have wildly different precision requirements. A reasonable candidate for the first practical vibrational problem is the calculation of zero-point energy or low-lying transitions in a vibrational problem to sub-chemical precision. See the SI for further discussion.

Detailed resource estimates are beyond the scope of this work, but these order-of-magnitude differences between vibrational and electronic Hamiltonians are noteworthy. These three metrics (term counts, locality, and $W$) seem to support the postulate that a vibrational problem instance may achieve quantum advantage before an electronic problem instance.

\begin{figure}
    \centering
    \includegraphics[width=0.99\linewidth]{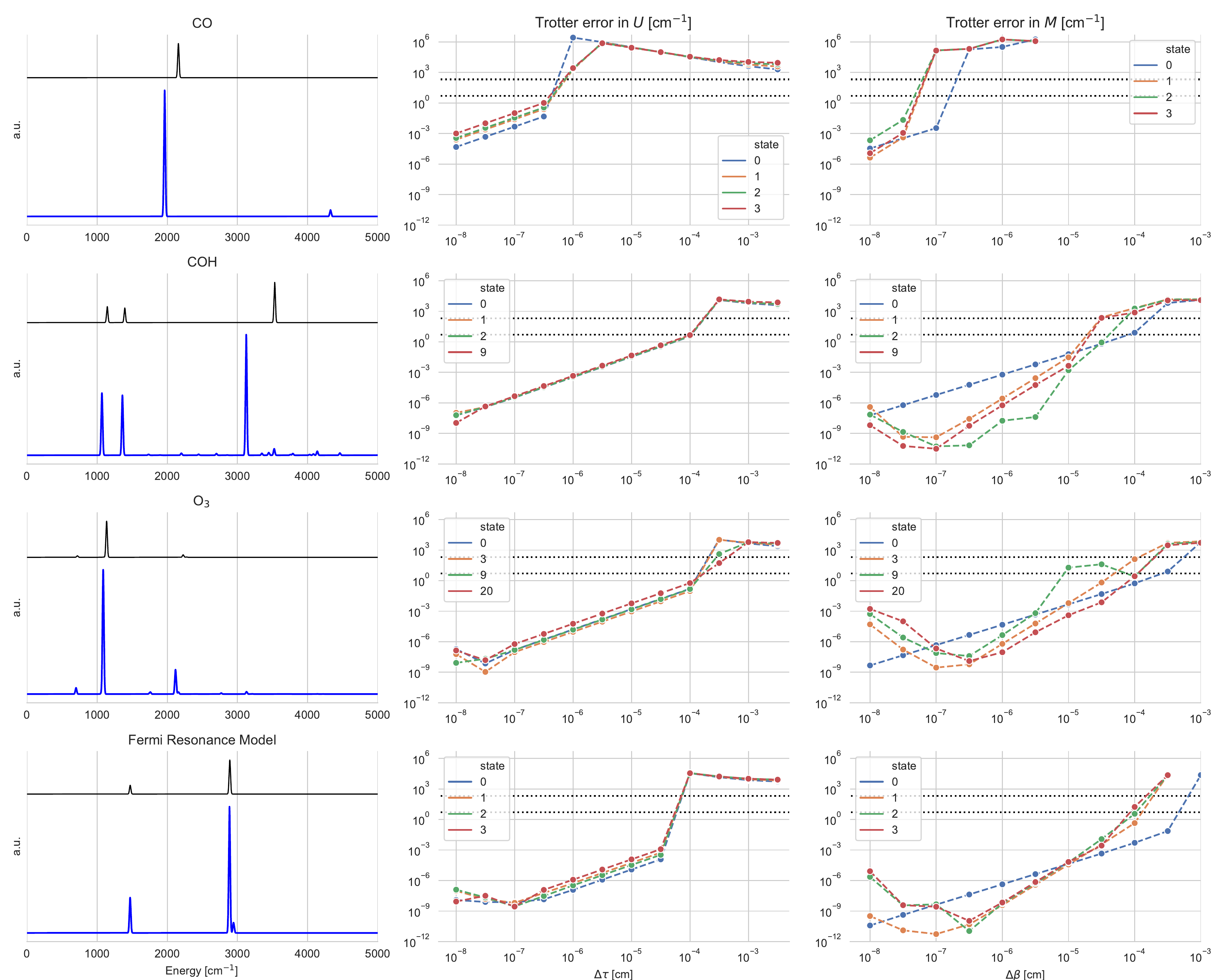}
    \caption{Vibrational infrared spectra for carbon monoxide (CO), the isoformyl radical (COH), ozone (O$_3$), and a Fermi resonsance model Hamiltonian. The first column shows the infrared spectra (blue) and their harmonic approximation (black) in arbitrary units, summing intensities in all Cartesian directions. Peaks were broadened with Gaussians of arbitrary standard deviation 10 \invcm. The second column shows Trotterization error in the quantum time propagator, which is relevant to long-term algorithms. The third column shows Trotterization error in the imaginary time evolution operator, relevant to some NISQ approaches. Excited states in the third column are found using the folded Hamiltonian method. Horizontal dotted lines are drawn at 5 \invcm, an arbitrary high-accuracy threshold required for many spectroscopy applications; and at 209 \invcm, equal to $k_B T$ at room temperature and approximately half of chemical accuracy.}
    \label{fig:spec_results}
\end{figure}

\textbf{IR spectra.} 
As a proof of concept, we performed numerical simulations and error analyses on four vibrational Hamiltonians: carbon monoxide (CO), the isoformyl radical (COH), ozone (O$_3$), and a model Hamiltonian of Fermi resonance (Figure \ref{fig:spec_results}). 
%
%
%
%

We studied Trotter error for both the approximate unitary used in QPE and the approximate imaginary time evolution (ITE) operator appropriate for some nearer-term algorithms. For the former case, we constructed the unitary matrix
\begin{equation}\label{eq:trott_qprop}
\tilde U(\Delta\tau) = \prod_k^{N_P} e^{-i \Delta\tau a_k P_k}.
\end{equation}
where $\Delta \tau$ is the time step. This is a first-order Trotter approximation to the quantum propagator $U(\Delta\tau)\equiv\exp(-i \Delta\tau H)$. We then diagonalized $\tilde U(\Delta\tau)$ and compared the ordered eigenvalues to the exact result. 



In our simulation of ITE, we instead constructed the operator
\begin{equation}\label{eq:trott_ite}
\tilde M(\Delta \beta) = \prod_k^{N_P} e^{- \Delta \beta a_k P_k}
\end{equation}
where $\Delta \beta$ is an ITE step. For all but the ground state, formula \eqref{eq:trott_ite} used the Pauli representation of the folded Hamiltonian, not of the original Hamiltonian. Folded Hamiltonians were used in order to highlight the use of a method that allows one to skip irrelevant eigenstates, effectively implementing spectral window focusing. 
Calculations were implemented using SciPy \cite{scipy2020}.

Figure 3 shows the infrared spectra (blue). The harmonic approximations (black) are plotted; the contrast between the two plots demonstrates the importance of including higher-order anharmonic terms that are hard to simulate classically. Qualitative differences such as the the extra peaks that appear (\textit{e.g.} at 2940 \invcm ~in the Fermi resonance Hamiltonian) tend to be difficult to obtain classically, often failing under perturbation theory \cite{sibert14_fermires,Sibert:2019_Review}.

The second column shows the Trotterization error in some of the more intense transitions' eigenvalues against increasing $\Delta \tau$. 
These are related to long-term algorithms, both in running QPE and in dynamical simulations. 
The third column approximates the ITE operator's error by Trotterization with finite length $\Delta \beta$. 
These are more relevant to noisy intermediate-scale quantum (NISQ) algorithms, both for ITE \cite{motta19_qite} and variational anstazae based on ITE \cite{mcardle19_ite}. 

Errors in $U$ are mostly independent of the eigenstates, while errors in $M$ are distributed over many orders of magnitude even for fixed $\Delta \beta$. This may be partly because each folded Hamiltonian is in fact a different  Hamiltonian.  Notably, all non-monotonic behavior in the ITE error plots arise only in folded Hamiltonians---the cause of this behavior is intriguing but unclear. CO requires the smallest (\textit{i.e.} worst) time step, which we hypothesize might be due to the lack of favorable error cancellation, as cancellation may be more prominent in Hamiltonians with more terms. There was no clear trend with respect to the (Froebenius) norms of the Hamiltonians (data not shown), though in the worse case (carbon monoxide) the error becomes unacceptably large approximately when $\Delta\tau$ and $\Delta\beta$ have order of magnitude comparable to the inverse norm of the Hamiltonian ($1/ \|H\|$).  The resulting error trends give an indication of the step sizes needed for accurate simulation of small molecules, though further study is needed to determine broadly applicable relationships between vibrational problem instances and Trotter error.



\section{Outlook}
\label{sec:concl}

Although molecular electronic structure is often the first candidate offered for near-term quantum simulation of a real-world substance, we have provided evidence suggesting that molecular vibrational structure will likely, for a given energy precision, achieve quantum advantage first. 
After considering previously unidentified requirements in designing quantum algorithms for vibrational spectra, we have presented approaches for solving this class of problems on both near-term and long-term quantum computers, addressing the components that make this mathematically distinct from the electronic structure problem:
calculating transition amplitudes with respect to a non-unitary operator and 
calculating high-lying excited states. 
Future research should focus on more detailed resource counts including estimates of circuit depth and gate complexity, as well as inclusion of rotational and other degrees of freedom. This work advances the applicability of quantum computation for atmospheric science, many biomolecular interactions, fuel combustion, gas-phase reactions, and astrochemistry, while implying that some focus for near-term quantum applications ought to shift to scientifically relevant vibrational problems.






%
%
%
%
%
%
%
%
%
%
%

\section*{Acknowledgments}
D.P.T. acknowledges support from Texas A\&M University startup funding and the Robert A. Welch Foundation (A-2049-20200401). Portions of this research were conducted with high performance research computing resources provided by Texas A\&M University HPRC. F.P. acknowledges support from the National Science Foundation through grant No. CHE-1453204. We thank Pauline J. Ollitrault for a discussion regarding excited state calculations and encodings, and we thank Jhonathan Romero for discussions regarding measurement counts for VQE. The authors declare no competing interest.

\section*{Supplementary Information}
Additional details regarding mapping vibrational modes to qubits, near-term quantum algorithms for finding spectra, the complexity comparison between vibrational and electronic problem classes, counting terms for fermionic and bosonic operators, and calculation of potential energy and dipole surfaces.



\bibliography{main}

\bibliographystyle{Science}

\end{document}



\baselineskip24pt


\maketitle



\tableofcontents


\section{Mapping vibrational modes to qubits}
\label{apx:bosonmapping}

The operator for a single $d$-level particle (included a truncated bosonic mode) may be expressed as
\begin{equation}
A = \sum_{l,l'=0}^{d-1} c_{l,l'} | l' \kb l |.
\end{equation}

In order to use a qubit-based quantum computer, each level must first be mapped to a bit representation, before the operator is mapped to a sum of products of Pauli matrices. For example, a 4-level particle with an operator $B = | 2 \kb 3 | + | 3 \kb 2 |$ would map to

\begin{equation}
\begin{split}
| 2 \kb 3 | + | 3 \kb 2 | \xmapsto{\textrm{Std. Binary}}& | 10 \kb 11 | + | 11 \kb 01 |   \\
=&   | 1 \kb 1 | \otimes | 0 \kb 1 |   +   | 1 \kb 0 | \otimes | 1 \kb 1 |    \\
=& \half \left ( \sx_0 - \sz_1 \sx_0    \right)
\end{split}
\end{equation}
where the least significant bit (qubit) is labelled 0. In the last step, the following identities are used:

\begin{equation}\label{eq:op01}
|0\ra\la 1| = \half ( \sx + i\sy ) 
\end{equation}
\begin{equation}\label{eq:op10}
|1\ra\la 0| = \half ( \sx - i\sy ) 
\end{equation}
\begin{equation}\label{eq:op00}
|0\kb 0| = \half (I + \sz),
\end{equation}
\begin{equation}\label{eq:op11}
|1\kb 1| = \half (I - \sz).
\end{equation}

In the case of vibrational (bosonic) degrees of freedom, one would truncate at a level of $d$ that preserves the accuracy one requires \cite{lee14_truncated,woods15_bosonic}. 
We use the Gray code for the numerics presented in this work; for a more thorough study of the choice and tradeoffs for different mappings, see reference \cite{sawaya20_dlev}.



\section{NISQ method for transition amplitudes}
\label{sec:ibe_trans}



Before summarizing the method for calculating arbitrary transition amplitudes, we give two known methods for calculating $| \la \eta_i | \eta_j \ra |^2$, an important primitive. In the first method, one implements $U_i^\dag U_j$. Thereafter, the fraction of measurements that equal the all-zero vector $|0\ra^{\otimes N_q}$ is equal to the overlap squared \cite{havlek19_qml}. This method does not require additional qubits, though the final circuit depth is equal to the sum of the two unitaries' depths. The second method is to used a SWAP test \cite{buhrman01_qfingerprint} or destructive SWAP test \cite{garcia13_destrswap}, which doubles the number of qubits but increases the depth only by a small constant factor.



The near-term algorithm for transition amplitudes $|\la i | \mudir | j \ra|^2$ is based on work by Ibe and coworkers \cite{ibe20_transition}.

The full expression for the transition amplitudes is
\begin{equation}
\begin{split}
|\la i |\mudir| j \ra|^2 =
\sum_k b_{k\alpha}^2 |\la i |R_{k\alpha}| j \ra|^2 \\
+\sum_{k<l} b_{k\alpha}b_{l\alpha} \Big[ 2|\la i |V_{kl,+}^\paralpha| j \ra|^2 + 2|\la i |V_{ij,-}^\paralpha| j \ra|^2 
- |\la i |R_{k\alpha}| j \ra|^2 - |\la i |R_{l\alpha}| j \ra|^2 - |\la i |R_{l\alpha}R_{k\alpha}| j \ra|^2  \Big].
\end{split}
\end{equation}
Terms $|\la i | V_{kl,\pm}^\paralpha | j \ra|^2$ are determined by preparing the state $U_i^\dag V_{kl,\pm}^\paralpha U_j|0\ra$. 
Terms such as $|\la i |R_{k\alpha}| j \ra|^2$ are determined by preparing state 

$$U_i^\dag \exp( i \frac{\pi}{2} R_{k\alpha}) U_j|0\ra$$
and counting zero strings.
Expressions $\exp(\pm i\theta R_{k\alpha})$ can be implemented in short depth using well known primitives \cite{mikeike11}. 
The number of required circuits scales quadratically with the number of Pauli strings $N_R^{\paralpha}$ in $\mudir$.

The advantage of Ibe et al.'s approach is that Hadamard tests \cite{mitarai19_newhadamtest} are not needed. Hadamard tests would require controlled-$U_i$ unitaries. Assuming the quantum hardware uses one- and two-qubit gates, this would often require each three-qubit gate to be decomposed into many one- and two-qubit gates \cite{mikeike11}, considerably increasing the circuit depth.

\section{Counting for fermionic operators}
\label{apx:count_elecstruct}



Here we summarize how the Pauli string counts were determined for molecular Hamiltonians. The electronic structure Hamiltonian takes the form

\begin{equation}
H_{ES} = \sum_{p\sigma q\tau} h_{p\sigma,q\tau} a^\dag_{p\sigma} a_{q\tau} + \sum_{pqrs \sigma\tau\mu\nu} h_{p\sigma,r\mu,s\nu,q\tau} a^\dag_{p\sigma} a^\dag_{r\mu} a_{s\nu} a_{q\tau}
\end{equation}
where Latin letters label spatial orbitals and Greek letters label the spin.

We assume that a real (as opposed to complex) basis is used. Fermionic commutation rules and spin orthogonality lead to the following symmetries \cite{helgaker13}.  First,

\begin{equation}
h_{PQRS} = h_{RSPQ}
\end{equation}
and
\begin{equation}
h_{PQRS} = h_{QPRS} = h_{PQSR} = h_{QPSR},
\end{equation}
which leads to an eight-fold symmetry. Including orthogonality of spin degrees of freedom leads to
\begin{equation}
h_{p\sigma,q\tau,r\mu,s\nu} = h_{pqrs}\delta_{\sigma\tau}\delta_{\mu\nu}.
\end{equation}

Finally, the following terms vanish:

\begin{equation}
\{a_i^\dag a_i^\dag a_j a_j , a_j^\dag a_j^\dag a_i a_i\} \rightarrow 0,
\end{equation}

\begin{equation}
a_i^\dag a_i^\dag a_i a_i \rightarrow 0,
\end{equation}

\begin{equation}
a_i^\dag a_i^\dag a_j a_k \rightarrow 0.
\end{equation}


For each category of fermionic Hamiltonian term, we now consider the number of Pauli strings resulting from the Jordan-Wigner mapping, though the Pauli string count (but not the locality) is the same for the Bravyi-Kitaev mapping. The Jordan-Wigner encoding maps fermionic degrees of freedom to qubits such that fermionic commutation relations are retained. The mapping is defined as 

\begin{equation}
\begin{split}
a_p^\dag \mapsto \left( \prod_{m<p} Z_m  \right)\sigma_p^+ \\
a_p^\dag \mapsto \left( \prod_{m<p} Z_m  \right)\sigma_p^- 
\end{split}
\end{equation}
where $\sigma^{\pm} \equiv \left( X \mp iY \right)/2$.

In our counting procedure, we avoid double-counting Pauli terms. For instance, terms like $Z_i$ appear in both one- and two-electron operators, but they are counted only once.

The one-electron terms lead to

\begin{equation}
a_i^\dag a_i \rightarrow Z+I = \half(I - Z_i)
\end{equation}
and 
\begin{equation}
a_i^\dag a_j + a_j^\dag a_i \rightarrow \half( X_i Z^{\bigotimes j-i-1} X_j + Y_i Z^{\bigotimes j-i-1} Y_j).  
\end{equation}


Non-vanishing two-orbital two-electron terms lead to

\begin{equation}
a_i^\dag a_j^\dag a_j a_i + a_j^\dag a_i^\dag a_i a_j \rightarrow \{I,Z_i,Z_j,Z_iZ_j\}
\end{equation}

Based on the above symmetries, non-vanishing three-orbital terms lead to
\begin{equation}
\begin{split}
\{a_i^\dag a_j^\dag a_k a_i,\cdots\}_{(4)} \cup \{a_i^\dag a_j^\dag a_i a_k,\cdots\}_{(4)}& \rightarrow \textrm{4 Pauli strings} \\
\textrm{Example: } \{a_0^\dag a_1^\dag a_2 a_0,\cdots\}_{(4)}& \cup \{a_0^\dag a_1^\dag a_0 a_2,\cdots\}_{(4)} \\
\rightarrow \{Z_0 X_2 X_3 ,  Z_0 Y_2 Y_3 ,  X_2 X_3 ,  Y_2 Y_3 \}
\end{split}
\end{equation}
where each set of four operators leads to the same set of Pauli strings. Subscripts denote the number of Pauli strings in the bracketed set.

Finally, consider four-orbital terms with 8-fold symmetry. One such set of terms leads to four Pauli strings:
\begin{equation}
\begin{split}
\{a_i^\dag a_k^\dag a_l a_j,\cdots\}_{(4)} \rightarrow \textrm{4 Pauli strings} \\
\textrm{Example: } \{a_1^\dag a_3^\dag a_5 a_7,...\}_{(4)} \rightarrow \{ &X_1 Z_2 X_3 Y_5 Z_6 Y_7, \\
 &X_1 Z_2 Y_3 Y_5 Z_6 X_7,  \\
 &Y_1 Z_2 X_3 X_5 Z_6 Y_7, \\
 &Y_1 Z_2 Y_3 X_5 Z_6 X_7 \}  )
\end{split}
\end{equation}




To numerically validate our manual counting procedure, we used OpenFermion \cite{openfermion} and Psi4 \cite{psi4} to calculate the number of Pauli strings required for the electronic structure problem of a collection of hydrogen atoms. Hydrogen atoms were placed on a cubic lattice with spacing 0.6 \AA, with a random perturbation in each direction drawn from a Gaussian of standard deviation 0.05 \AA. We used the minimal STO-3G basis, resulting in a number of qubits equal to 4 times the number of hydrogen atoms. The canonical orbitals used were determined from the Hartree-Fock calculation of Psi4. All Pauli string counts were within 10\% to 30\% of our manual analytical counts, a difference that we attribute primarily to the software truncating terms smaller than $10^{-6}$.

\section{Counting for bosonic operators}
\label{apx:count_bosonic}


A third-order vibrational Hamiltonian in normal coordinates has four types of terms: $p_i^2$, $q_i^2$, $q_i^3$, $q_i^2q_j$. A fourth-order Hamiltonian has eight types, with the inclusion of $q_i^4$, $q_i^3q_j$, $q_i^2q_j^2$, and $q_i^2q_jq_k$.




All of our mappings use the Gray code \cite{sawaya20_dlev}.
Here we provide the Pauli mappings or Pauli counts for the different types of many-body terms, for both $d$=4 and 8. These are used for counting the numer of terms in each type of Hamiltonian.



--- Harmonic: ---

\begin{equation}
\begin{split}
p_0^2 + q_0^2 &\xrightarrow{d=4} 4I - 1Z_0 Z_1 - 2Z_1 \\
&\xrightarrow{d=8} 8I - 1 Z_0 Z_1 Z_2
\end{split}
\end{equation}

--- 3rd-order: ---
\begin{equation}
\begin{split}
q_i^3 &\xrightarrow{d=4} \{ X_0, X_0 Z_1, Z_0 X_1, X_1 \} \\
&\xrightarrow{d=8} \textrm{(16 Pauli strings)}
\end{split}
\end{equation}

\begin{equation}
\begin{split}
q_i^2 q_j &\xrightarrow{d=4} \textrm{(20 Pauli strings)} \\
&\xrightarrow{d=8} \textrm{(144 Pauli strings)}
\end{split}
\end{equation}

--- 4th-order: ---
\begin{equation}
\begin{split}
q_i^4 &\xrightarrow{d=4} \textrm{(I \& 5 Pauli strings)} \\
&\xrightarrow{d=8} \textrm{(I \& 18 Pauli strings)}
\end{split}
\end{equation}

\begin{equation}
\begin{split}
q_i^3 q_j &\xrightarrow{d=4} \textrm{(16 Pauli strings)} \\
&\xrightarrow{d=8} \textrm{(192 Pauli strings)}
\end{split}
\end{equation}

\begin{equation}
\begin{split}
q_i^2q_j^2 &\xrightarrow{d=4} \textrm{(I \& 24 Pauli strings)} \\
&\xrightarrow{d=8} \textrm{(144 Pauli strings)}
\end{split}
\end{equation}

\begin{equation}
\begin{split}
q_i^2q_jq_k &\xrightarrow{d=4} \textrm{(80 Pauli strings)} \\
&\xrightarrow{d=8} \textrm{(1728 Pauli strings)}
\end{split}
\end{equation}

\begin{figure}
    \centering
    \includegraphics[width=0.95\linewidth]{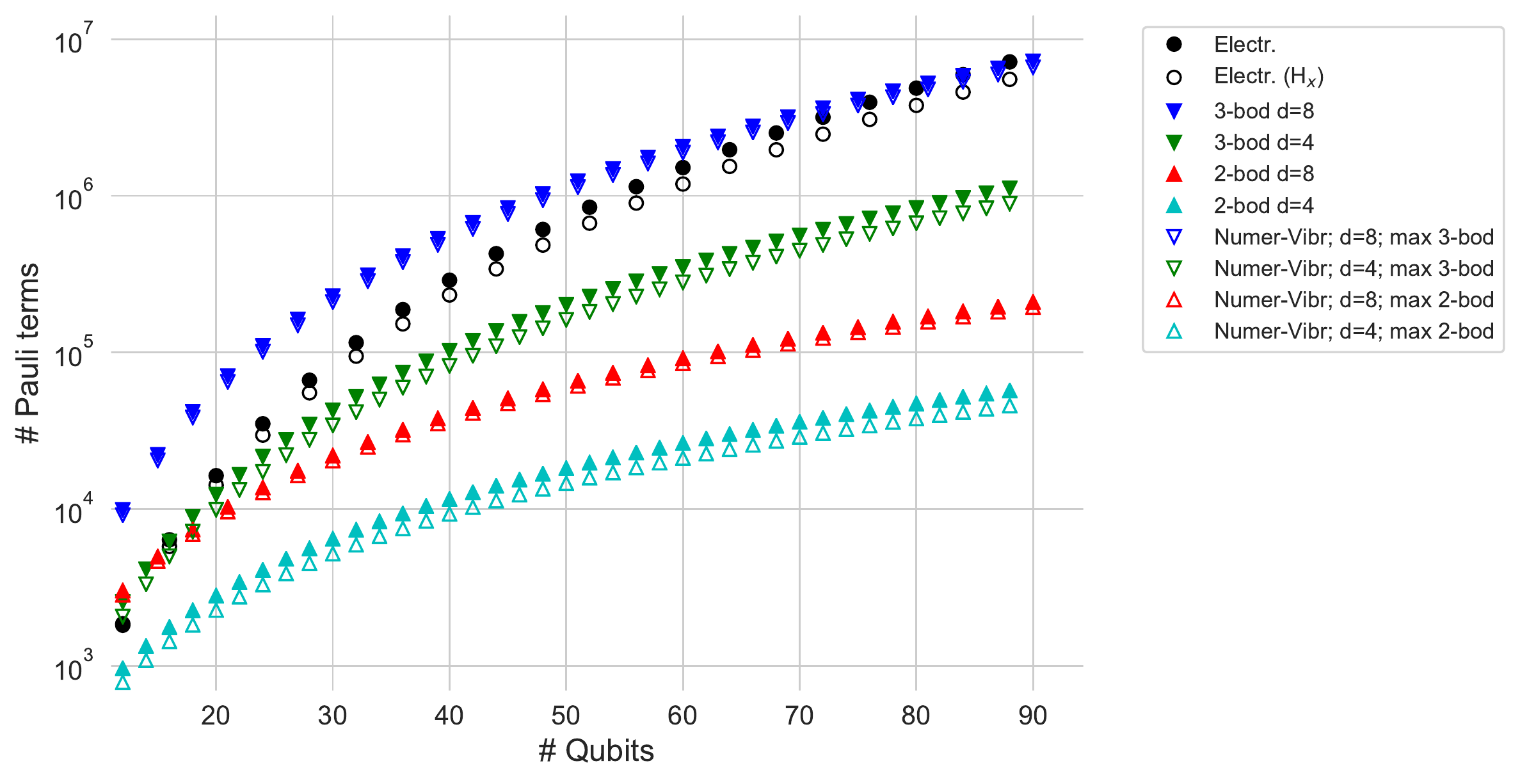}
    \caption{Pauli string counts, including both analytical and numerical (random vibrational Hamiltonian) results.}
    \label{fig:apx_all_term_counts}
\end{figure}

In order to numerically validate our counting procedure, we prepared vibrational Hamiltonians with randomized non-zero couplings for all possible terms up to fourth order. We omitted these ``numerical'' vibrational results from the main text in order to avoid over-crowding the figure. The analytical and numerical results for all Hamiltonians are shown in Figure \ref{fig:apx_all_term_counts}.

\section{Coefficients and precision}
\label{sec:meas}

\begin{table}[!t]
\begin{tabular}{|ll|cc|ccc|}
\hline
\multicolumn{1}{|c}{} & \multicolumn{1}{c|}{\# Qubits} & max ($\{|a_k|\}\setminus |a_I|$ )       & $W$                   & \multicolumn{3}{c|}{$10^{-6}\: W / \epsilon$}                                               \\
\multicolumn{1}{|c}{} & \multicolumn{1}{c|}{}          & (Ha)                 & (Ha)                  & $\epsilon$ = 100 cm$^{-1}$ & $\epsilon$ = 10 cm$^{-1}$ & $\epsilon$ = 1 cm$^{-1}$ \\
\multicolumn{1}{|c}{} & \multicolumn{1}{c|}{}          &                      &                       & 455 $\mu$Ha                    & 45.5 $\mu$Ha                  & 4.55 $\mu$Ha                 \\ \hline
Vibrational           & 24                             & 0.045                & 0.146                 & 0.321                       & 3.21                      & 32.1                     \\
2-body                & 36                             & 0.066                & 0.244                 & 0.536                       & 5.36                      & 53.6                     \\
$d=4$                 & 48                             & 0.088                & 0.358                 & 0.787                       & 7.87                      & 78.7                     \\ \hline
Vibrational           & 24                             & 0.045                & 0.187                 & 0.411                       & 4.11                      & 41.1                     \\
3-body                & 36                             & 0.066                & 0.387                 & 0.851                       & 8.51                      & 85.1                     \\
$d=4$                 & 48                             & 0.108                & 0.695                 & 1.53                       & 15.3                      & 153                      \\ \hline
Vibrational           & 24                             & 0.087                & 0.276                 & 0.607                      & 6.07                      & 60.7                     \\
2-body                & 36                             & 0.126                & 0.450                 & 0.989                      & 9.89                      & 98.9                     \\
$d=8$                 & 48                             & 0.166                & 0.651                 & 1.43                       & 14.3                      & 143                      \\ \hline
Vibrational           & 24                             & 0.087                & 0.336                 & 0.738                       & 7.38                      & 73.8                     \\
3-body                & 36                             & 0.126                & 0.675                 & 1.48                       & 14.8                      & 148                      \\
$d=8$                 & 48                             & 0.192                & 1.197                 & 2.63                       & 26.3                      & 263                      \\ \hline \hline
V$_2$ model           & 24                             & 0.471                & 2.41                  & 5.30                       & 53.0                      & 530                      \\
V$_3$ model           & 36                             & 0.639                & 3.35                  & 7.36                       & 73.6                      & 736                      \\
V$_4$ model           & 48                             & 0.752                & 4.72                  & 10.4                       & 104                       & 1040                     \\ \hline
Cr$_2$ model          & 24                             & 0.142                & 1.70                  & 3.74                       & 37.4                      & 374                      \\
Cr$_3$ model          & 36                             & 0.193                & 2.41                  & 5.30                       & 53.0                      & 530                      \\
Cr$_4$ model          & 48                             & 0.280                & 3.08                  & 6.77                       & 67.7                      & 677                      \\ \hline
Mn$_2$ model          & 24                             & 0.647                & 2.75                  & 6.04                       & 60.4                      & 604                      \\
Mn$_3$ model          & 36                             & 0.901                & 3.79                  & 8.33                       & 83.3                      & 833                      \\
Mn$_4$ model          & 48                             & 1.067                & 4.74                  & 10.4                       & 104                       & 1040                     \\ \hline
Fe$_2$ model          & 24                             & 1.38                & 5.47                 & 12.0                       & 120                       & 1200                     \\
Fe$_3$ model          & 36                             & 1.84                & 7.97                 & 17.5                       & 175                       & 1750                     \\
Fe$_4$ model          & 48                             & 1.73                & 10.1                 & 22.1                       & 221                       & 2210                     \\ \hline
FeMoco \cite{li18_chanfemoco}      & 154                            & 11.8                    &       &                            &                           &                          \\ \hline
\end{tabular}
\caption{Comparing the magnitudes of Hamiltonians. $W$ = $\sqrt{ \sum_{k \neq I} a_k^2 }$ is related to the Frobenius norm and also relates to the number of measurements required for VQE.}
\label{tbl:norms}
\end{table}


As discussed in the main text, for both near- and long-term Hamiltonian simulation the required computational resources depend partly on both the desired precision and some measure of the Hamiltonian's magnitude (such as a matrix norm). For instance, as Gonthier et al. recently pointed out \cite{gonthier20_meas}, it is not just the Pauli counts and lengths but also the coefficients of each term that are relevant to determining computational resource estimates. Their work mentioned this in the context of the measurements required for VQE, though it is true for most other Hamiltonian simulation techniques as well.

For a set of electronic and vibrational Hamiltonians described below, we calculate the following quantity (reproduced from the main text) as a way to compare matrix magnitudes:
\begin{equation}\label{eq:Wnorm}
W = \sqrt{ \sum_{k \neq I} a_k^2 },
\end{equation}
where ${k \neq I}$ signifies that the coefficient in front of the identity operator is excluded. $W2^{N_q/2}$ is an upper bound to the Frobenius norm, defined as $|H|_F=\sqrt{\sum b_{ij}^2}$, where $b_{ij}$ is element of the $i$th row and $j$th column of the Hamiltonian. We choose this metric because it is trivial to calculate and because it directly relates to resource estimates, such as the number of measurements required in VQE \cite{gonthier20_meas}.

It is beyond the scope of this work to provide actual resource estimates, as extensive calculations and the choice of specific molecules would be needed (indeed, entire studies have been dedicated estimating resources for a few molecules or a single molecule \cite{reiher17_femoco,li18_chanfemoco,elfving20_indust}). Instead, we have considered three metrics that are rough proxies for computational complexity: term counts, Pauli length, and $W$. Support for our postulate of vibrational quantum advantage comes from the fact that these approximations (which are pessimistic for the vibrational case) show approximately an order of magnitude smaller values (in term counts and in $W$), for problems larger than 40 qubits. 

As stated, we used pessimistic assumptions to construct vibrational Hamiltonians, partly in order to mitigate the fact the electronic structure calculations are approximate models. As before, we include up to either 2- or 3-body mode coupling terms and $d=\{4,8\}$. Harmonic modes were set to evenly spaced values from 1000 to 4000 \invcm. Every possible third-order term was given a coefficient of 400 \invcm ~and every possible fourth-order term given 40 \invcm. It is unlikely that a molecule would include such dense mode coupling, and the first vibrational problem to exhibit quantum advantage will by definition be a problem with lower coupling density.

For electronic structure, our goal was to obtain order-of-magnitude estimates for some chemical elements that might be the first to see quantum advantage. It is reasonable to speculate \cite{reiher17_femoco,elfving20_indust} that the first electronic structure simulation to show quantum advantage will be for a molecule containing transition metal elements, because the classical coupled cluster algorithm accurately treats many main group molecules (\textit{e.g.} hydrocarbons and water clusters) up to more than a thousand orbitals \cite{yoo10_ccstd_water,gyevi19_ccsdt_org,elfving20_indust}. CCSD(T) and other state-of-the-art methods often cannot accurately treat molecules with transition metal elements, which is part of the reason that quantum computational resource estimates to date have focused on such compounds \cite{reiher17_femoco,li18_chanfemoco,elfving20_indust}. Notably, the toy H$_x$ (hydrogen box) model we used to calculate term counts in fact showed much greater values of $W$ than those of the transition elements, though these are not particularly useful chemical systems and may not require a quantum computer to begin with.

We created simple electronic structure models of twelve transition metal molecules: V$_x$, Cr$_x$, Mn$_x$, Fe$_x$, where $x \in \{2,3,4\}$ and closed-shell anions were used for V$_3^-$ and Mn$_3^-$. All molecules were constrained to have singlet multiplicity. Using a minimal STO-3G basis, we ran OpenFermion \cite{openfermion} and Psi4 \cite{psi4} to determine the canonical orbitals. We then froze the canonical orbitals corresponding to atomic orbitals 1s$^2$ 2s$^2$ 2p$^6$ 3s$^2$ 3p$^6$. We chose an active space of 6 orbitals (12 spin-orbitals) per atom, the lowest-energy orbitals excluding the frozen core.

As is well known to chemists but perhaps less well-known in the physics community, different chemistry applications require dramatically different precisions. A qualitative understanding of an electronic structure problem may require a low accuracy of just 10 milliHartrees (mHa), thermodynamics observables (for both electronic and vibrational calculations) often require an accuracy of 1 mHa, and interpreting transition lines in a vibrational spectrum may require precisions as low as 5 $\mu$Ha.



Table \ref{tbl:norms} shows matrix magnitudes $W$ and the maximum non-identity Pauli coefficient for each Hamiltonian. 
$W$ ought to be compared only between Hamiltonians of the same qubit count. The far right columns in table \ref{tbl:norms} give the quotient $W/\epsilon$ for three different values of $\epsilon$. As different applications require different levels of precision, this quotient is useful because it provides a comparison between matrix magnitude and $\epsilon$. Time complexity for different Hamiltonian simulation algorithms are dependent on both $\epsilon$ and matrix magnitude \cite{berry15_hamsimoptimal,berry15_hamsimtaylor,childs18_pnas,low17_qsp,childs19_theorytrott}, though the relationship is non-trivial and normally uses a more standard metric of matrix magnitude such as the spectral norm.

As a sanity check we show the maximum Pauli coefficient of a published fermionic Hamiltonian for the FeMoco cofactor (Li \textit{et al.} \cite{li18_chanfemoco}), a widely studied biological molecule containing several transition metal elements. We include this value because their careful procedure for choosing the orbitals and active space is nearer to what one would do for a real quantum calculation. We are cautious not to make direct comparisons, due to the higher qubit counts (154 qubits) and presence of the fifth-row element molybdenum.

The first $\epsilon$ of 100 \invcm ~$\approx$ 0.455 mHa is just below ``chemical accuracy'' of 1.6 mHa, an energy precision that is acceptable for many applications. The stated goal of many electronic structure calculations, including above-mentioned work in quantum computational resource estimates, is to achieve sub-chemical accuracy. This is a reasonable goal for electronic structure calculations containing transition metals. Notably, $k_B T$ is $\sim$0.6 kcal/mol or $\sim$209 \invcm at room temperature, implying that thermal fluctuations in a typical laboratory are on the same order as this common standard for computational accuracy.

There will be molecules and complexes for which a highly accurate PES may be achieved using CCSD(T) or other methods, but for which the vibrational zero-point energy (ZPE) is classically intractable  \cite{harding17_combust,kapil19_anharmonicfree}. Calculating this ground-state energy (\textit{i.e.} ZPE) is a good first place to look for quantum advantage, because it might require no better than chemical accuracy (discussed below) for some applications. However, if a quantum device is capable of solving ZPE, the device is not far from being able to calculate low-lying transitions. Only a doubling of depth (or alternatively a doubling of qubits) would be required in order to use the variational quantum deflation method to obtain low-lying excited states and transitions \cite{higgott19_vqd}.

Notably, ``chemical accuracy'' of 1.6 mHa is not sufficient for some applications. For instance, because of the exponential behavior of the Arrhenius equation, rate constants may be incorrect by about an order of magnitude. Another example for which chemical accuracy is insufficient is vibrational eigenstates, between which the spacing is often much smaller than 100 \invcm.

The higher accuracy of 1 \invcm ~ $\approx$ 4.55 $\mu$Ha is sometimes referred to as spectroscopic accuracy, though the term does not have a universally accepted definition \cite{fortenberry13_qff,sharma14_spectracc,brown17_watermelt}. The important point is that the interpretation of vibrational spectra often requires a high accuracy of between 1 \invcm ~ and 10 \invcm. Notably, when a PES is being calculated using electronic structure, the accuracy of the PES cannot be worse than the accuracy required for the spectroscopic problem.

It is not just the precision and the norm or magnitude of the Hamiltonian, but also its locality that determines the required resources. Complexity theoretic results have shown that a more local Hamiltonian will require fewer resources \cite{berry15_hamsimoptimal,berry15_hamsimtaylor,childs18_pnas,low17_qsp,childs19_theorytrott}. Hence the advantage from the lower magnitude of the vibrational Hamiltonians will be improved even further relative to electronic Hamiltonians due to the lower locality, suggesting that focusing only on $W$ is again pessimistic against the vibrational case.

A comment on the number of measurements required in VQE is merited as well. Gonthier \textit{et al.} recently studied the number of measurements required for estimating the electronic energy of a set of organic molecules \cite{gonthier20_meas}. The quantity $W$ is closely related to the factor that Gonthier \textit{et al.} calculate to determine measurement counts. If one were to take the naive route of measuring only one Pauli term expectation value per circuit run, then $\big ( W/\epsilon \big )^2$ would be proportional to the number of measurements. (As a side note, we do not necessarily consider the number of measurements to be the limiting factor for VQE. We would argue that the primary concern for near-term hardware should be circuit depth, which would also be dependent on desired accuracy. Measurements in principle can be parallelized across devices; on the other hand, if the required circuit depth is too great for the hardware, there is little one can do.)

In practice, for the near-term VQE algorithm one desires as few circuit runs as possible. Gonthier \textit{et al.} report numerical results fitted to a simple power law, showing for a set of organic molecules that the number of required measurements growing as $\m O(N_q^{5.1})$ to $\m O(N_q^{5.8})$ when Pauli terms are grouped together, and growing as $\m O(N_q^{2.3})$ to $\m O(N_q^{3.6})$ when using the so-called basis rotation grouping \cite{huggins19_basisrot,gonthier20_meas}. (The latter method requires additional circuit depth that grows with the number of qubits, appended to the end of the state preparation ansatz.)

For vibrational Hamiltonians, on the other hand, the number of VQE circuit repetitions for two-body vibrational Hamiltonians is upper bounded by $\m O(N_q^1)$. This is because two-body terms can be measured simultaneously; one may measure terms on modes $((0,1),(2,3),\cdots)$ followed by $((0,2),(1,3),\cdots)$ and so on, where integer pairs correspond to mode pairs. Analogously, the number of circuit repetitions for three-body vibrational Hamiltonians scale at worst as $\m O(N_q^2)$. Hence both the lower values of $W$ and the apparently lower measurement scaling suggest that fewer measurements will be required for the vibrational case, for a given precision.

In summary, the three quantities of term count, locality, and $W$ suggest that vibrational problems will require fewer quantum resources for a given precision. As chemical accuracy (order of $\sim$ 1 mHa) is required for order-of-magnitude accuracy in rate constants (due to the Arrhenius equation \cite{cramer04}) regardless of which which type of degree of freedom is being simulated, we suggest that vibrational zero-point energy calculations and/or low-lying vibrational transitions are a reasonable place to look for early quantum advantage.

\section{Comparing Hamiltonians}
\label{apx:scaling}

We use this section to elaborate on some points made in the main text, regarding comparisons between vibrational and electronic problems.

Note that the 3rd- and 4th-order vibrational Taylor terms in $H_{\textrm{anharm}}$ (defined in the main text) in fact yield at most 2- and 3-body interactions, respectively, as every higher-order term necessarily includes $q_i^2$ in the product if one is using normal modes. 
This leads to either a $\mathcal O(M^2)$ or $\mathcal O(M^3)$ scaling in the number of terms for this subset of molecules, compared to $\mathcal O(N^4)$ for electronic structure of an arbitrary molecule, where $N$ is the number of spin-orbitals. Though we discuss high-lying excited states in our work, note that for this consideration of quantum advantage we are assuming that only low-lying vibrational states are being calculated. Very high-lying vibrational states may require larger truncations of vibrational modes or the use of more costly methods such as folded Hamiltonians.


An essential comment is merited regarding the direct comparison between these two problem classes. For the comparison between vibrational and electronic problems, the appropriate independent variable is indeed the number of qubits (not the number of electron orbitals or vibrational modes). This is because most of the earliest problem instances for which one would need a quantum computer are likely to be those for which exact diagonalization is required (because we are assuming that even methods such as perturbation theory fail). 
Hence to compare different problem types it is appropriate to make direct comparisons only between two problem instances with \textit{similarly sized Hilbert spaces}, which is the same as saying similar qubit counts. 
And as a practical matter, available quantum hardware will have a given qubit count and maximum circuit depth, and the question at hand would be whether that given hardware would be able to perform some classically intractable problem instance.

It is important to point out that one should assume that state-of-the-art methods will be used to reduce the number of qubits required to simulate \textit{whichever} problem is considered, leading to substantial reductions in the number of effective vibrational modes or electronic orbitals. In vibrational problems, one may choose a coordinate system that is efficient in terms of having low truncation requirements and fewer appreciable coupling terms \cite{wilson80,Sibert:2019_Review}. For electronic structure, one carefully chooses the active space of orbitals \cite{reiher17_femoco,li18_chanfemoco,vonburg20_catalysis}. The point is, the use of ``qubit reduction'' methods does not change the overall argument unless the modified problem changes one of the stated criteria, \textit{e.g.} if the problem is no longer well-described by a fourth-order Hamiltonian, in which case we exclude such a problem from the discussion anyway. Methods for reducing qubit count in electronic structure should still produce a Hamiltonian with a similar number of terms and with similar magnitude. The appropriate comparison to make is still between two Hamiltonians of similar qubit counts. Further, it may be the case that some qubit reduction techniques would work equally well for bosonic and fermionic problems.


Note that several encoding choices are possible. Instead of using the Gray code, one may choose to use the unary encoding (also called direct mapping or one-hot encoding) with vibrational degrees of freedom, which would tend to reduce circuit depth at the expense of increasing the number of qubits \cite{mcardle19_qvibr,sawaya20_dlev}. We note that if we had used the unary encoding, the Hamiltonians would have been even simpler (fewer terms and shorter Pauli lengths, and similar matrix magnitude $W$). In principle this simpler Hamiltonian would either strengthen our conclusion (because the Hamiltonian can be simulated in shorter depth) or be a moot case (if the unary mapping requires too many qubits for whatever device characteristics you happen to be assuming). 

Finally, we stress again that our argument holds for only some subset of vibrational problems; many vibrational problems would require a \textit{e.g.} sixth-order Hamiltonian for accurate simulation, and some 
vibrational problems require more coupling terms than others.


\section{Potential energy and dipole surfaces}
\label{sec:cfour}

The potential energy surface and dipole derivatives for CO and COH were calculated at the CCSD(T)/ANO1 level with the CFOUR package, version 2.1 \cite{cfour}. We used the package's documented scripts for calculating anharmonic frequencies by finite difference. Energies were calculated in parallel to obtain the quadratic, cubic, and quartic (including three-body) force constants. 


All energy units are \invcm. The resulting fourth-order Hamiltonian for carbon monoxide is

\begin{equation}
H_{CO} = 2157.96 \frac{q^2 + p^2}{2} -736.66 ~q^3  + 210.97 ~q^4.
\end{equation}

The fourth-order Hamiltonian for the isoformyl radical is

\begin{equation}
\begin{split}
H_{COH} &= 1143.24 \frac{q_1^2 + p_1^2}{2} \\
        &+ 1393.46 \frac{q_2^2 + p_2^2}{2} \\
        &+ 3530.65 \frac{q_3^2 + p_3^2}{2} \\ 
        &+ -16.83 ~q_1q_1q_1
        + 51.76 ~q_1q_1q_2 \\ 
        &+ 40.02 ~q_1q_2q_2
        + 87.05 ~q_2q_2q_2 \\ 
        &+ 413.74 ~q_1q_1q_3
        -116.13 ~q_1q_2q_3 \\ 
        &-35.26 ~q_2q_2q_3
        -92.65 ~q_1q_3q_3 \\ 
        &-119.29 ~q_2q_3q_3
        -489.34 ~q_3q_3q_3 \\ 
        &+ 22.83 ~q_1q_1q_1q_1
        -10.84 ~q_1q_1q_1q_2 \\ 
        &-10.48 ~q_1q_1q_1q_3
        -0.49 ~q_1q_1q_2q_2 \\ 
        &-40.20 ~q_1q_1q_2q_3
        -252.07 ~q_1q_1q_3q_3 \\ 
        &+ 7.00 ~q_1q_2q_2q_2
        -3.37 ~q_1q_2q_2q_3 \\ 
        &+ 6.96 ~q_2q_2q_2q_2
        -6.15 ~q_2q_2q_2q_3 \\ 
        &-17.44 ~q_2q_2q_3q_3
        + 26.32 ~q_1q_2q_3q_3 \\ 
        &-3.13 ~q_1q_3q_3q_3
        -4.33 ~q_2q_3q_3q_3 \\ 
        &+ 66.64 ~q_3q_3q_3q_3,
\end{split}
\end{equation}
where $q_1$ is the bending mode, $q_2$ is the CO stretch, and $q_3$ is the OH stretch.

The first-order transition dipoles were found to be

\begin{equation}
\begin{split}
\mu^{(x)} \sim -0.33854  ~q_1 -0.268687 ~q_2  -0.011334 ~q_3  \\
\mu^{(y)} \sim -0.057874 ~q_1 -0.023912 ~q_2 + 0.175823 ~q_3 \\
\end{split}
\end{equation}
where the equilibrium dipole is irrelevant because it cancels out due to orthogonality. Units are in Debye, though we considered only relative intensities in our calculations.







For ozone (isotope $^{16}$O$_3$) we used previously published PES \cite{Barbe74_o3pes} and DMSs \cite{Adlergolden_1985_o3dps}. 





Our model Hamiltonian for Fermi resonances has two vibrational modes, taking the form 


\begin{equation}
\hat H_{FR} = \omega_0 \left( a_0^\dag a_0 + \frac{1}{2} \right) + \omega_1 \left( a_1^\dag a_1 + \frac{1}{2} \right) + h_{001} q_0^2 q_1 
\end{equation}


We choose frequencies and couplings that are typical for the bending and stretching modes of two CH stretches within a methyl (CH$_3$) functional group, a well-known example of Fermi resonance. We set $\omega_0$ = 1470 \invcm (bend), $\omega_1=2890$ \invcm  (stretch), and $h_{001}=30$ \invcm \cite{sibert14_fermires}. A necessary condition for Fermi resonance is that $\omega_1 \approx 2\omega_0$, which is met here as $\omega_1/\omega_0 \approx 1.966$. In our model calculations, for the first-order Taylor terms of dipole moment surface $\mu$ we set $m_0=m_1$. The Fermi resonance leads to the extra peak at $\sim$2940 \invcm \ appearing in Figure 3 of the main text, a peak which is not present in the harmonic approximation.

\bibliography{main}

\bibliographystyle{Science}